\documentstyle[11pt,aaspptwo]{article}

\def\beq{\begin{equation}}
\def\eeq{\end{equation}}
\def\beqar{\begin{eqnarray}}
\def\eeqar{\end{eqnarray}}

\def\la{\mathrel{\mathpalette\fun <}}
\def\ga{\mathrel{\mathpalette\fun >}}
\def\fun#1#2{\lower3.6pt\vbox{\baselineskip0pt\lineskip.9pt
  \ialign{$\mathsurround=0pt#1\hfil##\hfil$\crcr#2\crcr\sim\crcr}}}

\def\pd{\partial}
\def\msol{\hbox{$M_\odot$}}
\def\emax{\hbox{$E_{\rm max}$}}
\def\vmax{\hbox{$v_{\rm max}$}}

\def\li#1{\hbox{${}^{#1}$Li}}
\def\be#1{\hbox{${}^{#1}$Be}}
\def\b1#1{\hbox{${}^{1#1}$B}}
\def\c1#1{\hbox{${}^{1#1}$C}}
\def\o1#1{\hbox{${}^{1#1}$O}}
\def\ne2#1{\hbox{${}^{2#1}$Ne}}
\def\ti4#1{\hbox{${}^{4#1}$Ti}}

\begin{document}

\title{Nuclear and Gamma-ray Production by Supernova Ejecta}

\author{Brian D. Fields,$^{1,2}$ 
Michel Cass\'{e},$^{3}$ 
Elisabeth Vangioni-Flam,$^1$
and Ken'ichi Nomoto${}^4$}

\begin{center}
${}^1${\it Institut d'Astrophysique \\
98 bis, boulevard Arago, Paris 75014, France}
\medskip \\
${}^1${\it Department of Physics,
University of Notre Dame \\
Notre Dame, IN 46556 USA}
\medskip \\
${}^2${\it CE-Saclay,
DSM/DAPNIA/Service d'Astrophysique \\
91191 Gif sur Yvette cedex, France}
\medskip \\
${}^3${\it Department of Astronomy 
and Research Center for the Early Universe \\
University of Tokyo, \ Tokyo, Japan}
\end{center}

\begin{abstract}
The fastest ejecta of supernova explosions propagate as a precursor to
the main supernova shock wave, and can be quite energetic.  The
spectrum of such fast ejecta is estimated based on recent analytic and
numerical supernova models, and found to be a power law having a
cutoff at an energy of order 10 MeV/nucl, the precise value of which
depends on the supernova mass and energetics.  With cutoffs in this
range there can be significant flux with energies above the thresholds
for $\gamma$-ray and Li, Be, and B production.  These nuclear
interactions should inevitably accompany the passage of prompt
supernova ejecta through the surrounding medium.  The fast particle
composition is that of the outermost layers of the progenitor; if the
progenitor experienced significant loss of its envelope to a companion
or by mass loss, then the composition is nonsolar, and in particular,
metal-rich.  Such a composition and spectrum of fast particles is
required to explain the recent COMPTEL observations of $\gamma$-ray
emission in Orion.  Supernova ejecta from one progenitor having lost a
large amount of mass are shown to quantitatively account for the Orion
$\gamma$ rays, and to imply the presence of other, weaker lines and of
a supernova explosion in Orion in the last $\la 10^4$ years.
Implications for this mechanism as a mode of nucleosynthesis, and as a
potential supernova diagnostic, are discussed.  Model dependences and
uncertainties are noted, and the need is shown for accurate
measurements of $\gamma$-ray and spallation production cross sections
at energies near threshold.
\end{abstract}


\section{Introduction}

An exploding supernova is one of the most powerful engines in the
Galaxy.  Modern supernova models predict that the ejecta should have a
distribution in velocity, with the outermost layers being ejected at
great speed.  Indeed, it was originally hoped (e.g., Colgate \&
Johnson \cite{cj}) that these energetic particles might be the cosmic
rays, and thus supernovae would be the direct agents of cosmic ray
acceleration.  This has been shown to be incorrect (e.g., Cass\'e \&
Goret \cite{cg}; Meyer \cite{meya,meyb}), and instead it is now
believed that the supernova blast leads indirectly to cosmic ray
production via shock acceleration (as reviewed, e.g., in Blanford \&
Eichler
\cite{be}, and Jones \& Ellison \cite{ell}).  

Nevertheless, very fast ejecta are indeed created by supernova
explosions.  Indeed, some have energies above thresholds for nuclear
interactions, in particular for $\gamma$-ray and Li, Be, \& B (LiBeB)
production.  These interactions should inevitably accompany the
passage of fast ejecta through the medium surrounding the supernova.
Given the presence of these interactions, the crucial question is
whether they are numerous enough to be detectable.  Indeed, if the
flux of the fast particles is sufficiently high, the $\gamma$-rays
could offer a new and unique window on the explosion, and the LiBeB
could provide a record of such events, integrated over the history of
the Galaxy.

In this paper we estimate the flux of the fast supernova ejecta and
find that it is in fact sufficiently energetic and intense to lead to
significant $\gamma$-ray production, and less significant LiBeB
production.  It is thus plausible that this mechanism is of great
interest as a diagnostic of supernovae in general.  More particularly,
it is interesting in light of recent Orion $\gamma$-ray observations,
which can be accounted for by prompt ejecta a single type Ic supernova
(if it indeed arises from a CO core progenitor as recently suggested
by Nomoto et al.\ \cite{nomoto}).

\section{Composition and spectrum of energetic massive star ejecta}
\label{sec:compspec}

Consider a supernova exploding in some medium, e.g.\ the ISM or, in
the case of Orion, a molecular cloud.  The ejecta will interact with
this medium, which may have been altered locally by the supernova
progenitor.  For example, if there was a strong wind, it will have
cleared out a cavity in the local medium in which the local material
is replaced by that of the wind.  Moreover, a typical progenitor
begins life as an O star whose UV radiation partially clears out the
region near the star.  In fact, it is only important for our purposes
that the explosion be in {\it some} medium, and we will see that the
particles we are interested in will travel much further than the
lengthscale of local inhomogeneities.

After the explosion the bulk of the material will form a shock and
pass through the usual phases (free-coasting, Sedov, etc.).  But the
precursors to the blast, the fastest particles ejected, are dilute and
will not participate in the shock.  They will instead propagate out
and begin to lose energy predominantly through ionization losses,
eventually to be stopped.  These are the particles in which we are
interested.  Specifically, we want to calculate the intensity,
spectrum, composition, and duration of the precursor nuclei.

Before making detailed calculations, let us fix the order of magnitude
of the effect.  Typical supernova velocity spectra have maximum
velocities of the order of 40,000 km/s, i.e., with particle energies
(per nucleon) up to 8 MeV/nucl.  Typical $\gamma$-ray thresholds (for
reactions between an $\alpha$ particle and a heavy nucleus) are $\sim
2$ MeV/nucl, and spallation thresholds a few times higher; proton
thresholds are higher still.  A threshold energy of $E_{\rm th} = 2$
MeV/nucl corresponds to a velocity of $v_{\rm th} =$ 20~000 km/s; the
amount of mass ejected above this energy is roughly $M(v>v_{\rm th})
\sim 0.001 \msol$, i.e., a tenth of a percent of the total ejecta.
This material has a kinetic energy of order 
$4 \times 10^{48}$ erg, i.e., about 0.4\% of the supernova
energy.  This is to be compared with a typical efficiency of $\sim 1\%
$ for acceleration of galactic cosmic ray supernova shocks (e.g.,
Gaisser \cite{gais}).  Since prompt ejecta have comparable energetics,
it is plausible that their effects could be important.

We are thus interested in calculating the initial spectrum of
particles and determining their nuclear interactions as they propagate
away {}from the explosion.  The initial spectrum is calculated as
follows.  Imagine as well that the star emits at total number $N_i$ of
isotopes $i$.  The spectrum is obtained from the calculations of the
mass, velocity, and composition of supernova shells.  Specifically, we
know the velocity $v(M)$ and mass fraction $X_i(M)$ as a function of
mass coordinate $M$.  We want the number spectrum $\pd N_i/\pd E$ for
each species $i$ of interest.  We have, for the total number of
particles,
\beq
\label{eq:mcount}
N_i = \int_{M_{\ell}}^{M_{\rm h}} \frac{dM_i}{A_i m_{\rm p}}
    = \int_{M_{\ell}}^{M_{\rm h}} dM \ \frac{X_i}{A_i m_{\rm p}}
\eeq
integrated over the ejected shells mass between the inner 
$M_{\ell}$ and outer $M_{\rm h}$ mass coordinates of the ejecta.
But by definition we also have
\beqar
\label{eq:ndef}
\nonumber
N_i & \equiv & \int_{E_{\ell}}^{E_{\rm h}} dE \ \frac{\pd N_i}{\pd E} \\
\nonumber
    & = & \int_{v_{\ell}}^{v_{\rm h}} dv \ \frac{dE}{dv}
                                          \ \frac{\pd N_i}{\pd E} \\
    & = & \int_{M_{\ell}}^{M_{\rm h}} dM \ \frac{dv}{dM} 
                          \ \frac{dE}{dv} \ \frac{\pd N_i}{\pd E}
\eeqar
and so by equations (\ref{eq:mcount}) and (\ref{eq:ndef}) we have
\beq
\frac{\pd N_i}{\pd E} = \frac{X_i}{A_i} \ 
      \left( m_{\rm p} \; \frac{dv}{dM} \; \frac{dE}{dv} \right)^{-1}
\eeq
and in the nonrelativistic case which holds here, we may simplify
further to
\beq
\frac{\pd N_i}{\pd E} = \frac{1}{m_{\rm p}^2} \ \frac{X_i}{A_i} \ 
	   \frac{1}{v} \, \frac{dM}{dv} 
\eeq

Now we may define and then estimate
the rate of LiBeB or $\gamma$-ray production
by this flux.
Formally, the rate is given by
\beqar
\label{eq:rate}
\frac{dN_l}{dt} & = & \sum_{ij} \ n_j 
        \int_{E_{\rm i}}^\infty dE \ \sigma_{ij}^l \; \frac{\pd N_i}{\pd E}
        \; v \\
\nonumber
     & = & \sum_{ij} \ n_j \; \frac{1}{A_i m_{\rm p}^2} \
            \int_{E_{\rm i}}^\infty dE \ \sigma_{ij}^l \; X_i \;
            \frac{dM}{dv}
\eeqar
Note that the spectrum $\pd N_i/\pd E$ will be out of equilibrium
and so time dependent (in contrast to the steady state of
galactic cosmic rays).
Because of the energy losses, the fast particle flux
will be continuously degraded in intensity and energy
over time.  For the purposes of a first approximation, 
however, imagine a constant
flux lasting for a duration $\Delta t$.  The appropriateness of this
approximation will be discussed in \S \ref{sec:prop}

{}From eq.\ (\ref{eq:rate}) it is clear that what is crucial to our
calculation is the nature of the fastest ejecta, i.e., the composition
$X_i$ and the velocity spectrum $dM/dv$ of the outermost supernova
shells.  Both of these will be sensitive to the progenitor mass and
history.  Specifically, the outermost composition (and indeed the
pre-explosion mass) will depend on the amount of mass loss the
progenitor has suffered.  Although the outermost layers will be born
with the composition of the ISM at the time, with sufficient mass loss
this could be shed and the inner nucleosynthetic shells can be
exposed.  Thus mass loss considerations are crucial for determining
the fast particle composition.

Since mass loss can change the progenitor mass considerably, it also
affects the velocity of the ejecta, and thus influences the
normalization of the velocity spectrum.  However, it appears that the
shape of the velocity spectrum is generically a power law to a
reasonable approximation.  In a beautiful analytic treatment,
Nadyozhin (\cite{nad}) has examined the passage of the type II
supernova shock through the outermost layers.  He shows that in this a
self-similar solution holds, and calculates the velocity spectrum of
the fastest layers.  He finds the spectrum to have the form of a power
law with a cutoff at a maximum velocity \vmax.  Specifically, he finds
\beq
\label{eq:spec}
M(v) = \left\{ \begin{array}{cc}
		{\cal B} (v^{-s} - \vmax^{-s}) + \Delta M \ , 
                      & v < v_{\rm max} \\
		0 \ , & v \ge \vmax
              \end{array} \right.
\eeq
where ${\cal B}$ is a constant that depends on the details of
the hydrodynamics, while \vmax\ scales roughly as 
$\vmax \propto \sqrt{U/M_{\rm ej}}$
with the supernova mechanical energy
$U$ and the ejected mass $M_{\rm ej}$.\footnote{
Note there is an additional shell ejected at the highest velocity, but 
its mass is so small ($\Delta M \simeq 10^{-6} \msol$) 
that we ignore it.}
For SN 1987A,
Nadyozhin calculates
$\vmax = (26-40) \times 10^8$ cm/s, corresponding to an energy 
(per nucleon) cutoff of $\emax = 6-8$ MeV/nucl.  

As we will see in detail below, the presence of the cutoff velocity
\vmax\ is crucial in determining the nuclear interactions of the fast
particles.  Physically, the cutoff arises from the quenching of the
shock's acceleration of the outer layers of the supernova.  As the
shock moves through the decreasing density of the outer atmosphere,
the material is accelerated progressively faster in reaction to the
large radiation pressure of the shock.  This process ceases, however,
when the shock approaches the surface and the optical depth for the
shock radiation becomes comparable to the optical depth to reach the
photosphere.  At this point the radiation escapes from the putative
supernova as an ultraviolet flash, and the shock causes no further
direct acceleration (for further discussion see Nadyozhin \cite{nad}).

A very nice feature of the Nadyozhin spectrum is that it takes the
exceedingly simple form of a power law.  Thus one can immediately
deduce that the effective particle spectrum (cf.\ eq.\ \ref{eq:rate})
is also a power law $dM/dv \propto v^{-(s+1)}$, or
\beq
\frac{dM}{dv}(E) \propto E^{-(s+1)/2} \ , \ \ E < \emax
\eeq
For SN 1987A, Nadyozhin estimates that $s = 7.2$, which gives an
energy dependence of $E^{-4.1}$.  This result is supported by Ensman
\& Burrows (\cite{eb}) whose numerical model of SN 1987A was
specifically designed to calculate shock breakout.

The slope $s$ of the velocity spectrum (eq.\ \ref{eq:spec}) is related
to the hydrodynamic properties of the self-similar shock solution.  In
particular, $s = (n+1)/\gamma$, where $n$ is a power law index for the
density dependence on the distance to the surface, and $\lambda$ is a
similar index for the velocity (related to $n$ and the adiabatic
index).  At any rate, the universality of the power law solution is
very fortuitous, as it allows us to make general remarks about all
SNe.

The other major feature of the velocity spectrum which is equally
crucial for our purposes is the existence of the cutoff in velocity,
and thus in energy.  The cutoff is quite near to the $\gamma$-ray and
spallation thresholds, and so will greatly emphasize the differences
among the different thresholds.  To be sure, there is the
above-mentioned scaling with supernova mass and energy, so a lighter
pre-supernova star could perhaps achieve higher energies.
Regardlessly, the low energy behavior of the cross sections will be
crucial.  In particular, protons reactions are disallowed below $\sim
20$ MeV/nucl for C and O reactions, but since $\alpha$ thresholds are
lower by a factor $\sim 4$ in energy {\it per nucleon}, the components
of the ejecta in $\alpha$ particles, and heavier nuclei which can
interact with interstellar He, will be crucial.

The smallness of the cutoff energy means that the numerical
calculations are difficult to do correctly.  Results are very
sensitive to the generally unmeasured threshold behavior of the cross
sections.  Semianalytic fits to experimental data (Read \& Viola
\cite{rv}) do extrapolate down to threshold, but clearly there is a
need to understand this low-energy physics exceedingly well, and we
urge experimentalists to consider these measurements.  Indeed if
accurate cross sections were available, one could turn the problem
around, and use the $\gamma$-ray and spallation yields as a powerful
diagnostic on the fast ejecta's spectrum and composition.

Another word of caution: Nadyozhin's calculation was plainly not
intended for the use we make of it.  In particular, it is doubtful
that the real cutoff behavior is as sharp as he calculates it to be
(eq.\ \ref{eq:spec}).  In actual supernovae it is likely that the
quenching of the shock is a continuous process and so while the cutoff
scale may be correct, the details of the spectrum at these highest
energies may be quite complicated.  One should thus regard numerical
results obtained via eq.\ (\ref{eq:spec}) as rough indicators of the
yields one expects from these fast ejecta.  Nevertheless, we forge
ahead in applying our method to the recent $\gamma$-ray observations
of Orion.

\section{Gamma-ray Observations of Orion}
\label{sec:data}

The $\gamma$-ray production in the Orion complex is huge, as Bloemen
et al.\ (\cite{blo}) recently determined with the COMPTEL experiment
on the {\it Compton Gamma Ray Observatory}.  The most surprising
feature of the $\gamma$ emission from Orion is that the large flux is
confined to the $3-7$ MeV range.  Bloemen et al.\ (\cite{blo})
measured the flux in the 3-7 MeV band to be
\beq
\Phi_\gamma (3-7 \ {\rm MeV}) = \left( 1.0 \pm 0.15 \right) \times 10^{-4} \
    \ {\rm cm}^{-2} \ {\rm s^{-1}} \ \ .
\eeq
In this range, the spectrum shows features consistent with
de-excitation emission from the 4.44 MeV excited state of \c12${}^*$
and the 6.13 MeV state of \o16${}^*$; these features have roughly the
same amplitude.

The features at 4 and 6 MeV appear wide, which would have implications
for the nature of the accelerated particles due to the kinematics of
the nuclear excitations.  Fast C and O nuclei give rise to fast
\c12${}^*$ and \o16${}^*$ which have Doppler-broadened decay lines,
whereas fast protons and $\alpha$ nuclei lead to slow \c12${}^*$ and
\o16${}^*$ having narrow lines.  However, Bloemen et al.\ (\cite{blo})
caution that their lineshapes are not corrected for broadening due to
finite detector response, and to the widths are artificial, a point
also noted by Reeves \& Prantzos (\cite{rp}) and emphasized via
detailed calculations by Ramaty, Kozlovsky, \& Lingenfelter
(\cite{rkla,rklb}).

The COMPTEL experiment also set limits to the flux outside of the
$3-7$ MeV band; in particular, the flux in the $1-3$ MeV range was
constrained to be $\la 1/7$ that of the $3-7$ MeV range. The lack of
such lines is a crucial indicator of the lack of fast protons or
$\alpha$ particles, which would excite ambient Fe, Si, and Ne, leading
to emission in this range.

It is assumed that these lines have their origin in energetic particle
interactions with ambient matter in the Orion complex, but it is clear
that any explanation of the $\gamma$ emission will have to be
unconventional.  No lines are observed below the $3-7$ MeV band, and
no excess $\pi^0$ decays have been observed at higher energies (i.e.,
no excess over the level expected from galactic cosmic rays; see
Digel, Hunter, \& Mukherjee \cite{dhm}).  The if galactic cosmic rays
produced the $3-7$ MeV lines, they would also produce 1--3 MeV
$\gamma$-ray lines which are not observed; also the GCR intensity
would have to be much larger, which would give a high excess of
$\pi^0$ decay photons, again unobserved.  Moreover, the energetic
demand would be enormous.  This process is therefore excluded.

The most economic way to obtain the observed $\gamma$-ray flux is
through accelerated C and O nuclei, and above all, a suppression of
protons (Ramaty, Kozlovsky, \& Lingenfelter \cite{rkla,rklb};
Vangioni-Flam et al.\ \cite{elisa}; Cass\'e, Lehoucq, \& Vangioni-Flam
\cite{clv}; and Cass\'e et al.\ \cite{cvlo}).
Indeed, if one assumes merely that the Orion $\gamma$-rays arise from
energetic particle interactions, then one may deduce in a
model-independent way the particles must be a large flux with low
energy, predominantly composed of C and O rather than protons and
$\alpha$ particles. 

A low energy CO flux was suggested by the Bloemen et al.\ (\cite{blo})
paper itself, and the low energy character was assumed in all
subsequent efforts at explaining the $\gamma$-ray emission.  Bykov \&
Bloemen (\cite{bb}) and Nath \& Biermann (\cite{nb}) connect the
observation to theories of cosmic ray origin.  Bykov \& Bloemen
(\cite{bb}) propose a specific spectrum arising from collisions
between supernova shocks and stellar winds.  They use pure supernova
compositions in their calculations, but it remains unclear how much of
an admixture of normal material will be present in the hot regions
they consider.  Nath \& Biermann (\cite{nb}) favor acceleration in
shocks of winds of hot O and B stars; they argue that this naturally
gives the desired composition.  Ramaty, Kozlovsky, \& Lingenfelter
(\cite{rkla}) pointed out that while the observed $\gamma$-ray
features appear broad, there is insufficient experimental resolution
to exclude a narrow line component.  Parameterizing a low energy
spectrum, they tried different source compositions, showing that a
composition like that of a WC supernova wind would be needed to meet
the $\gamma$ spectrum and energetic constraints.  These authors have
also suggested (Ramaty, Kozlovsky, \& Lingenfelter
\cite{rklb}) that the low energy cosmic rays
could have an origin akin to that of the so-called anomalous component
of the cosmic rays observed at earth (a suggestion also made by Jin \&
Clayton \cite{jc}).

A low energy cosmic ray flux has implications for the nucleosynthesis
of Li, Be, and B (LiBeB), as first pointed out by Meneguzzi \& Reeves
(\cite{mrb}).  Cass\'e, Lehoucq, \& Vangioni-Flam (\cite{clv})
investigated the implications of such a process for LiBeB chemical
evolution.  They show that the LiBeB synthesis can be significant (for
an exposure time of $10^5$ yr), comparable to that of galactic cosmic
rays (e.g.\ Walker, Mathews, \& Viola \cite{wmv}).

Much of the work on the Orion $\gamma$-rays thus far has either: (1)
assumed arbitrary flux shape, or (2) assumed SN composition.  These
first approaches adopt a spectrum of the form $\propto E^{-s}$, with
$s = 0$ below an energy $E_{\rm c}$, and $n= 5-10$ above $E_{\rm c}$
(a form chosen for exploratory purposes, following Meneguzzi \& Reeves
(\cite{mrb}) and Ramaty, Kozlovsky, \& Lingenfelter (\cite{rkl79}))
and try different supernova compositions, as Bykov \& Bloemen
(\cite{bb}).  We propose a more specific scenario related to a type Ic
supernova explosion in a molecular cloud.

Our mechanism naturally has all of the needed features, provided only
that the supernova exploding within Orion be sufficiently massive to
have undergone significant mass loss.  Then the progenitor would be
stripped to a CO core, which would be ejected at high speeds (for a
fixed explosion energy, a smaller core mass has less gravitational
binding energy and so leads to higher kinetic energy for the ejecta).
Furthermore, neither the fast particle spectrum nor the number of
particles is arbitrary, but rather these are given by the supernova
model.  Thus the predictions of this process are specific and so the
mechanism is testable.

\section{Confrontation with the Orion Observation}

We have seen that theory allows us to compute $\dot{N_\gamma}$, while
the Orion observation report the flux $\Phi_\gamma$ at Earth.
These quantities are simply related by
\beqar
\nonumber
\label{eq:nobs}
\frac{d N_\gamma}{dt} & = & 4 \pi \; R^2 \; \Phi_\gamma \\
 & = & \left( 3.0 \pm 0.45 \right) \times 10^{39} 
        \ {\rm s}^{-1}
\eeqar
using $R_{\rm Orion} \simeq 500 \; {\rm pc}$ as the distance to Orion,
and assuming the flux to be emitted isotropically so that
the total subtended solid angle is $4\pi$.
Of course it is conceivable that the emission is anisotropic
due to, e.g., magnetic fields having somehow
focussed (or defocussed) the beam in our direction.  
In the lack of clear evidence for such effects, we will henceforth
make the simplest assumption regarding the angular distribution of the
radiation and consider it to be isotropic. 

To get a feel for the numerical results, let us
make an order of magnitude estimate of the $\gamma$-ray
emissivity.  The particles of interest are the ejecta
having a velocity larger than the reaction thresholds;
these particles have mass $M_{\rm ej}(v>v_{\rm th})$. 
Taking an energy ($E_{\rm min} = 2$ MeV/nucl) 
slightly above the threshold for gamma rays ($\sim$ 1.5 MeV
for $\alpha$+C), we have 
\beqar
\nonumber
\label{eq:est}
\dot{N_\gamma} & \sim & n_{\rm Orion} \, v \, \sigma_{\rm pC}^\gamma \,
      N_{\rm CO}(v>v_{\rm min}) \\
\nonumber
 & = &  n_{\rm Orion} \, v \, \sigma_{\rm pC}^\gamma
          \ \frac{X_i^{CO}}{A_i} \ 
          \ \frac{M_{\rm ej}(v>v_{\rm min})}{m_{\rm p}} \\
  & = & 1.0 \times 10^{39} \ {\rm s}^{-1} \ X_{\rm C} \\
\nonumber
  & & \hphantom{1.0} \times \
        \frac{n_{\rm Orion}}{20 \ {\rm cm}^{-3}} \ 
        \frac{\sigma_{\rm pC}^\gamma}{10 \ {\rm mb}} \
	\frac{M_{\rm ej}(v>v_{\rm min})}{0.03 \ \msol}
\eeqar
where we have taken the cross section to be a typical one in the low
energy range, about 10\% of the maximum, and have allowed for an
average density in Orion that is moderately enhanced over that of
the ISM.

Note that eq.\ (\ref{eq:est}) shows that for the fiducial values we
have chosen, the emissivity is just at the observed levels for the C
lines (which contribute about half of the total emissivity) but only
if $X_{\rm C}$ (and similarly $X_{\rm O}$) is close to unity.  That
this, the fast ejecta must be predominantly composed of C and O.  But
the fastest ejecta arises from material at the outermost layers of the
supernova progenitor.  One expects a ``normal'' supernova progenitor,
one not having significant mass loss, to have a composition similar to
that of the local ISM at the time of the progenitor's birth.  This
would give C and O mass fractions $< 10^{-2}$, which leads, via eq.\
(\ref{eq:est}), to a significant underproduction of $\c12^*$ and
$\o16^*$ lines.

However, if the progenitor has suffered significant mass loss, or
has lost its outer envelope to a companion star, then it is possible
that the inner nucleosynthetic shells could be left at the outer
layers of the progenitor.  In this case the fastest ejecta would have
a very nonsolar composition.  In particular, if the progenitor
is reduced to a CO core, then it would have 
$X_{\rm C} \simeq X_{\rm O} \simeq 0.5$; this is just the composition
required to explain the Orion observations.  
While these nuclei dominate, there remain some $\alpha$ particles
and also a small amout of Ne (having the form \ne22 rather than
\ne20).

We thus suggest that the Orion $\gamma$-rays are the result
of an explosion of a CO core supernova, similar to that
observed recently in the type Ic supernova 1994I 
(Schmidt, Challis, \& Kirchner \cite{sck}; Clocchiatti, Brotherton,
Harkness, \& Wheeler \cite{cbhw}).  
Nomoto et al.\ (\cite{nomoto}) argued that the observed
characteristics of this event are best understood in the
context of a CO model arising from envelope loss to a companion
(as opposed to mass loss of a single star).  
While such events are not the most common fates of massive stars,
they are not so unlikely:  Nomoto et al.\ estimate a ratio
of type Ic to type II explosions of $\sim 10$\%.

We now apply our formalism in more detail to the particular case of a
CO core explosion.  We use the results from the supernova model of
Nomoto et al.\ (\cite{nomoto}) (specifically the $v(M)$ and $X_i(M)$
relations) to generate the spectrum shown in Figure \ref{fig:coflux}.
As expected, most of the particles indeed lie at low energies, and the
composition varies over energies reflecting the variation of the
composition with mass shell.  Since the reaction thresholds are all
$\ge 2$ MeV/nucl, the only part of the flux relevant to our problem is
that at and above this scale.  At these energies, the composition does
not vary with radius (being that of the outmost shells).  Moreover,
the spectrum has, fortunately, settled down to a fairly smooth shape
close to a power law, with a slope of approximately $\sim 4$, in
reasonable agreement with the Nadyozhin analytical calculation.
Preliminary calculations using the velocity profiles for Wolf-Rayet
supernovae (Woosley, Langer, \& Weaver \cite{wlw}) give a similar
slope and so give further credence to the analytic solutions.

To investigate the effect of the flux, it is crucial to know its
behavior for energies above those specified in the Nomoto calculation.
We have chosen to extrapolate the flux to higher energies by taking a
power law with the slope of the last points.  Specifically, if we
write $\phi \sim E^{-\alpha}$, we find $\alpha \sim$ 3.5 around 1 MeV,
steepening to $\alpha \sim 4$ around 2 MeV; it is this last slope we
used in the extrapolation.

Following Nadyozhin, we wish to introduce a maximum energy $\emax$ to
the flux.  The self-similar solution provides the scaling mentioned
above, $\emax \propto (U/M_{\rm ej})^{1/2}$, which Nadyozhin
normalizes to SN1987A hydrodynamic models.  Scaling his result to our
CO core explosion, and assuming a total energy $U=10^{51}$ erg, gives
$\emax \sim 40$ MeV/nucl.  However, since the details of determining
\emax\ are difficult, the supernova energy is uncertain as well, we
will take \emax to be a parameter and we will investigate values in
the range $8 \ {\rm MeV/nucl} \le \emax \le 50 \ {\rm MeV/nucl}$.

Using this extrapolated flux we have evaluated numerically the
$\c12{}^*$, $\o16{}^*$, $\ne22{}^*$, and LiBeB production.  We assume
that the escape length $\Lambda_{\rm esc}$ is large and so the LiBeB
produced are all either thermalized through ionization losses (the
dominant process), or lost through inelastic collisions.  In fact,
since the ionization losses dominate, the results are insensitive to
the escape length.  Further we assume that the medium has a solar
system composition as given by Anders \& Grevesse (\cite{ag}).

Given these parameters, the 
ratios among the $\gamma$ and spallation rates are fixed;
however, to compute absolute rates requires
specification of the
Orion H density $n_{\rm H}$.  
Specifically, we find, for the total $\gamma$
production from $\c12{}^*+\o16{}^*$, we have
\beq
\frac{dN_\gamma}{dt} = 2.4 \times 10^{38} 
     \ \frac{n_{\rm H}}{1 \ {\rm cm}^{-3}}
     \ \rm{photons} \ \rm{s}^{-1} \ \ {}
\eeq
with very little variation over the chosen
range of \emax.
Comparing this to the measured value (eq.\ \ref{eq:nobs}), we see that
our mechanism can account for the observed Orion $\gamma$-ray flux
if the ejecta encounter an average density
\beq
\label{eq:nfix}
n_{\rm H} \simeq 12 \ {\rm cm}^{-3} \ \ {}.
\eeq
This is to be compared with to the estimated {\it average}
density of Orion, $n_{\rm H}^{\rm Orion} \simeq 5 \ {\rm cm}^{-3}$
(Goudis \cite{goudis}),  
the uncertainty in which is large enough so that it is in good
agreement with eq.\ (\ref{eq:nfix}).
We note, however, that
the density requirement rises with the time since the explosion,
since the flux is constantly decreasing in intensity.  As discussed 
below (\S \ref{sec:prop}) the timescale for significant loss 
is energy-dependent but at least of order 1 kyr.

Figure \ref{fig:emiss}(a) shows calculated the emissivity from excited
\c12, \o16, as well as
\ne22, plotted as a function of \emax.  The 
normalization is fixed by requiring the total $\c12{}^*+\o16{}^*$
emissivity to match the Orion observations (eq.\ \ref{eq:nobs}); as
discussed in the previous paragraph, this fixes a value of $n_{\rm H}$
for each \emax, which appears in Figure \ref{fig:emiss}(b).  We see
that over the preferred range of \emax, the $\gamma$-line ratios, and
the required density, show almost no variation, and what little change
there is has settled down by 10 MeV/nucl.  We find ${\rm C}^*/{\rm
O}^* = 2.7$, consistent with the Bloemen et al.\ (\cite{blo})
observations.

It is insufficient to show that the flux can produce the observed
$3-7$ MeV lines; we must also show there will not be lines outside
this range.  In particular, we must show that the $1-3$ MeV lines are
below the COMPTEL limits (c.f.\ \S\ref{sec:data}).  Here again, the
composition of the stripped supernova core proves advantageous.  In
particular, not only are the fast ejecta dominant in C and O, as
discussed above, but they also have relatively small abundances of
heavier elements.  In particular, the next most abundant element,
after C, O, and He, is \ne22.  Although its abundance is not large, it
is much larger than its solar proportion, indeed much larger than the
solar proportion of heavy elements.  Thus this will be the most
important contributor of heavy element lines, even more than those
caused by the $\alpha$ reactions in the ISM.

Indeed, we find $({\rm C}^*+{\rm O}^*)/{\rm Ne}^* \ga 13$ for all
values of \emax, which falls safely below the observational limits,
which are of order 7.  However, if ours is the mechanism for making
the C and O $\gamma$-rays, then there should be ${}^{22}$Ne lines
present at a level just below the current limits.  While COMPTEL may
not have a good chance to see them, INTEGRAL will see them in detail.
This will then provide a clean discrimination between our idea and
others, as our prediction of \ne22 lines (at 1.3 MeV), rather than
\ne20 lines (at 1.6 MeV), is likely to be unique.

Given the roughness of our estimate, and its closeness to the
observational limit, it is clear that the $1-3$ MeV window offers an
important constraint.  More detailed modeling is needed and is under
way (Ramaty, Vangioni-Flam, Fields, \& Cass\'e
\cite{rvfc}).
Indeed, we see that the generic sensitivity of the $\gamma$-ray
emission to the supernova composition and velocity structure makes the
$\gamma$ spectrum a powerful diagnostic tool.  Since all supernovae
should have fast ejecta, one could in general turn the problem around
and use the constraint as a probe of supernova ejecta.

Although a full study of fast particle interactions from all possible
supernovae is beyond the scope of this paper, nevertheless some
discussion of model dependence is in order.  We are interested in the
sensitivity to progenitor mass and mass loss, as manifested in the
flux intensity, spectrum, and composition.  To begin to address this
question we have studied variants to the supernova model we have used
thus far.

The hydrodynamics of the explosion determines the intensity and
spectrum, and the Nadyozhin (\cite{nad}) model suggests (eq.\
\ref{eq:spec}) that the spectral slope is fairly independent of the
initial mass, but the velocity scaling and energy cutoff depends on
$\vmax \propto \sqrt{U/M_{\rm ej}}$.  The scaling with mass means that
significant mass loss can be important since, for a fixed explosion
energy, it reduces the gravitational potential to be overcome and so
increases the final kinetic energy.  To test the accuracy of the
expected scaling with the explosion energy (at a fixed mass), we
compare the spectrum for a $2.1 \msol$ core at energies of
$(0.6,0.8,1)\times 10^{51}$ erg (Nomoto et al.\ \cite{nom95a,nom95b}).
Results appear in figure \ref{fig:escal}; we see that while the
scaling is not exact, it is a good approximation.  Note that changes
in energy scaling do not affect the intensity of the scaled flux, but
the shifts in the velocity scale change the intensity of the relevant,
highest energy flux.

To test the accuracy of the expected scaling with progenitor mass (for
a fixed explosion energy), we compare the spectrum from the $2.1
\msol$ CO core we have considered so far with that of CO cores with
masses $1.8$ (Iwamoto et al.\ \cite{iwa}) and $2.9 \msol$ (Nomoto et
al.\ \cite{nom95a}).  The results appear in figure
\ref{fig:masscal}.  Here we find the agreement is not as good;
indeed it appears that the particle fluxes are more similar without
the scaling.\footnote{Note, however, that the plot only demonstrates
the scaling of the overall intensity; the models on which the
calculations are based do not include the physics of the shock
breakout and so cannot show the energy cutoff.}  While for our
calculations we have used the numerical results anyway, it would be
interesting if the results were less sensitive to ejecta mass than
expected, as this would lead to larger fluxes (above threshold) than
expected.  At any rate we find that as a whole, the slope, the energy
cutoff, and the intensity is well-described by the very simple
Nadyozhin model itself coming from well-understood physics.

We now turn to the composition, which depends strongly on the mass
loss, and so is tied to much less certain physics.  Indeed, mass loss
is poorly understood and simply treated parametrically.  We have tried
using different mass loss rates; results appear in Table
\ref{tab:shell} At any rate, it is apparent from figure
\ref{fig:coflux}, the composition of the fastest shells is fairly
homogeneous.  Thus one need only determine the (uncertain) outermost
composition without having to get the full profile and convolving it
with the hydrodynamics.

\section{Propagation and Energetics}
\label{sec:prop}

Recall that in calculating the rates for nuclear interactions of the
fast particles we have assumed the flux to be present at full strength
for some time interval $\Delta t$, during which the spectrum is
constant.  While this is adequate for estimates of emissivity, it is
too simplistic.  At the energies of interest the particles are stopped
predominantly by ionization energy losses.  As these losses are a
strong function of energy and of $A$ and $Z$ (scaling as $Z^2/A$), the
particle spectrum and composition will change with time.

The calculation is done in detail in Fields (\cite{fcv}).  The basic
effect is that the spectrum continually becomes less intense, but also
harder.  To see this, imagine an (unphysical) spectrum that is a
$\delta$-function in energy.  It would degrade from its initial energy
according to the usual Bethe-Bloch $\partial E/\partial t$ formula.
Thus for a continuous spectrum, one expects the entire flux to degrade
in energy.  But since $\partial E/\partial t \propto E^{-1}$, lower
energy particles lose energy faster, so the spectrum, though less
intense. has an increasing average energy per particle.

An important related point is that the stopping ranges are larger than
size of Orion.  This means that if the fast particles are still in
Orion then the supernova is young.  More quantitatively, the range of
a 2 MeV/nucl oxygen nucleus in Orion is about 50 pc; it traverses this
distance in about 2 kyr; this is about a factor 3 smaller than the
crossing time, but it is also a lower bound; for particles above about
5 MeV/nucl the range is larger than Orion.  Thus we can use the
crossing time to put an approximate upper limit to the supernova date
of $\sim 10^4$ yr ago.  Of course, an accurate calculation must
include the fact that the exposure time is not the total stopping time
but rather the time it takes the fast particles to pass below the
threshold for $\gamma$-ray production.  Since the stopping time is a
strong function of energy ($t \propto E^2$) this difference will only
be important for the particles very close to the threshold energy.
This effect is discussed in Fields (\cite{fcv}).

The possibility of a relatively recent supernova is constrained by
observations of short-lived radioactivities.  In particular, the lack
of observed \ti44 ($\tau = 78.2$ yr) sets a lower limit to the
supernova age of order $\tau$, assuming that there is significant
\ti44 production in the supernova in question.  In fact, we can make
this point more quantitative.  The CO core supernova models we have
been using give \ti44 yields of order $M_{44} = 5 \times 10^{-5} \msol
- 2 \times 10^{-4} \msol$ (Nomoto et al.\ \cite{nom91}; Kumagai et
al.\ \cite{kum}; Thielemann et al.\ \cite{f-kt96}), a value close to
the $10^{-4} \msol$ inferred from observations in Cas A (Iyudin et
al.\ \cite{ti44}).  If the time since the explosion is $\Delta t$,
there will be a \ti44 emissivity of
\beq
\dot{N}_{44} = \frac{1}{\tau} \ \frac{M_{44}}{44 \; m_{\rm p}} 
               \ e^{-\Delta t/\tau}
\eeq
at an energy $E_\gamma = 1.16$ MeV.  The observational limit on such
lines in Orion is $L \la 3\times 10^{38} \ {\rm photons} \  {\rm s}^{-1}$ 
(namely the limit for the 1--3 MeV band), which gives
\beq
\label{eq:ti44}
\Delta t \ga \tau \ln \frac{M_{44}}{44 m_{\rm p} \tau L} 
  =  570 \ {\rm yr} \ \ .
\eeq
This is a strong constraint, being only a factor of 4 shorter than
the minimum particle stopping time.  
Taking this result at face value, we are led to infer
a possible age range of 0.6--10 kyr for the supernova.

However, caution is merited in using the \ti44 lower bound on the age.
The \ti44 yield is very uncertain, as it requires an accurate
knowledge of which of the deepest layers of the star are ejected, and
which fall back into the neutron star (i.e., the mass cut).  Different
mass cut choices are allowed and can lead to smaller \ti44 yields
(although the age limit depends only logarithmically on the mass
yield).  Also, we note that the observational limit we have used is
that of COMPTEL and is more suited for broad lines; Bloemen et al.\
(\cite{blo}) caution that narrow lines may be left undetected.  And in
any case, if the age is closer to the $\sim 10^4$ yr crossing time,
then there is no \ti44 to be seen by either COMPTEL or INTEGRAL.  On
the other hand, to take more optimistic viewpoint and turn the problem
around: if there has been a very recent supernova in Orion, the \ti44
lines (or lack thereof) can give important information about the mass
cut.

Finally, we turn to the question of energy requirements.  As pointed
out by Ramaty, Kozlovsky, \& Lingenfelter (\cite{rkla,rklb}), the
significant low-energy flux in Orion is a challenge to explain in a
way that is energetically economical.  These authors note that a flux
lasting long enough ($\ga 10^5$ yr) to produce significant amounts of
Li, Be, and B must be composed of relatively energetic particles ($\ga
30$ MeV/nucl) to avoid an unreasonable energy demand.

Our mechanism also must be shown to have reasonable energy
requirements.  To see most simply that indeed the energetics are
acceptable, recall that our particles have a very definite source: one
C+O core supernova.  Thus in this model the total energy required to
accelerate the particles and thus create the observed $\gamma$-rays is
exactly the (mechanical) energy of the supernova: $10^{51}$ erg.  We
note again that the flux intensity and composition of the particles is
fully determined by the supernova model, generally and in the specific
scenario we have adopted for Orion.  The energy is always fixed to be
that of the supernova, and so as a consequence we are {\it not}\ free
to adjust the number of fast particles; thus the energy budget is
automatically that of one supernova, but the resulting flux may or may
not be observable (depending notably on the ejected mass).

The number of $\gamma$-rays per unit energy injected in particles has
been calculated by other groups and so provides a useful quantity for
comparison of the results.  In our case, the total energy injected in
fast particles is above 2 MeV/nucl is $W = 1.4 \times 10^{50}$ erg,
about 14\% of the total energy (this is larger than the generic
estimate given in the beginning of \S \ref{sec:compspec} due to the
unusually low ejecta mass). The total number of $\gamma$-rays made in
Orion is the product of the present total emissivity and the
irradiation time: $N_\gamma = \dot{N}_\gamma \tau \simeq 10^{50}$
photons (using a conservative exposure time of $\tau = 1$ kyr).  This
gives an efficiency $N_\gamma/W \simeq 0.7$ photon/erg.  By way of
comparision, Ramaty, Kozlovsky, \& Lingenfelter (\cite{rkla,rklb})
find values around 10 photons/erg, using a spectrum with a
characteristic energy of 30 MeV/nucl, which is in fact more efficient.

The implications in this difference in efficiencies can be understood
by noting that the fast particle propagation is at all not the same in
the two models.  We do not use a steady state flux given by a thick
target solution, but rater an impulsive flux.  As a result, in our
case the exposure time is fixed as we have just discussed, and is in
fact much smaller that that of Ramaty et al.\ (\cite{rkla,rklb}) or
Cass\'e, Lehoucq, \& Vangioni-Flam (\cite{clv}).  As the flux has a
shorter duration its total energy requirements are less; on the other
hand, it also produces less LiBeB, as we now see.

\section{Li, Be, and B Production}

With the above discussion one can compute the time-integrated
production rates.  The formalism and discussion is found in Fields
(\cite{fcv}).  The most important point is that the nuclear
interaction rate proportional to $n_{\rm ISM}$ (cf.\ eq.\
\ref{eq:rate}), but the exposure is set by the stopping time which is
proportional to ${n_{\rm ISM}}^{-1}$.  Thus the production at a given
value of energy is related only to the stopping power; specifically,
it is proportional to the ionization range as expressed in g ${\rm
cm}^{-2}$.  Roughly speaking, one integrates over ionization range
rather than over time, to arrive at a yield that is a given number or
mass of LiBeB per supernova.

Thus the absolute yields of LiBeB are independent of the density of
the circumstellar medium; they are only dependent on the composition
of the ISM.  Consequently, for a fixed composition, the yield for a
given supernova is independent of the region in which it explodes.
However, while the same argument holds for the total number of
$\gamma$-rays, their {\it intensity} does depend on $n$, as we have
discussed above (cf.\ eq.\ \ref{eq:nfix}).  Note also that the
composition of the medium, and of the ejecta, is crucial.

We have calculated the LiBeB yields for the ISM with a solar system
composition.  In figure \ref{fig:mout} we plot the total mass of LiBeB
isotopes produced as a function of \emax.  We see that, unlike the
results for $\gamma$-ray emission, the LiBeB production is very
sensitive to \emax.  This may be understood very simply in terms of
cross section thresholds.  Typically, for spallation processes one has
$E_{\rm th}(\alpha+{\rm N}) < E_{\rm th}(\alpha+{\rm C}) < E_{\rm
th}(\alpha+{\rm O})$, so the low \emax\ behavior arises {}from the
$\alpha+{\rm N}$ reactions, and the rapid variations happen when
\emax\ rises above the $\alpha$ thresholds.

The implications of this mechanism for chemical evolution of LiBeB are
hard to foresee given the yields for just one supernova type (and a
special one at that).  Nevertheless, it is of interest to compare to
other suggested LiBeB production mechanisms in supernovae, e.g.\ the
$\nu$-process yields (Woosley \& Weaver \cite{ww})\footnote{ In fact,
other mechanisms of supernova production of LiBeB have been suggested,
most notably that of Colgate (\cite{colg}) and Dearborn et al.\
(\cite{dear}).  Both of these involve synthesis within the shock
itself, as opposed to our treatment of the fast particles' interaction
with the ISM.  Each has also been shown to have apparently fatal
problems, by Weaver (\cite{weaver}) and Brown et al.\ (\cite{brown}),
respectively.}.  These are a function of supernova mass and
metallicity, but at solar metallicity
\b11 production lies in the range 
$M_{\rm B} \simeq (1-30) \times 10^{-5} \msol$.  It has been shown
(Olive et al.\ \cite{opsv}; Timmes, Woosley, \& Weaver
\cite{tww}) that these yields, when combined with the
usual GCR production of LiBeB, are sufficient to explain the solar
\b11/\b10 ratio (but not \be9).  Thus it would seem that since the
$\nu$-process yields are significantly higher than that for our
particular supernova, it is unlikely that our mechanism can produce
the solar boron.  A more detailed analysis of the relative LiBeB
contributions of these processes as well as GCR will appear in a
forthcoming paper (Vangioni-Flam et al.\ \cite{elisa}).

While the supernova precursors appear to have trouble explaining the
solar B abundances, it remains however to be seen if they may explain
the Pop II abundances.  A useful diagnostic is the B/O ratio.  For the
CO supernova we have considered, the O production is $M_{\rm O} \simeq
0.2-0.4 \msol$ (Thielemann et al.\ \cite{f-kt94,f-kt96}).  This gives,
for high \emax, B/O $\simeq 3 \times 10^{-7}$ by number.  Comparing
this with the solar ratio (B/O)$_\odot = 1.2 \times 10^{-6}$, we see
that even if all of the solar O were produced by CO supernovae, the
most favorable assumption, we still find B production to fall short by
a factor of 2--3, again with realistic LiBeB production giving smaller
numbers.  However, in Population II, we have B/O $\sim 0.1 \; {\rm
(B/O)}_\odot \sim (3-6) \times 10^{-7}$.  Thus we would require only
$\sim 5-10\% $ of Pop II supernovae to be of this type in order to
have significant Pop II B production by our mechanism; this is to be
compared to the supernova type Ic/II ratio of $\sim 10\%$ (Nomoto et
al.\ \cite{nomoto}).

This suggests that the Pop II B (as well as Li and Be) may have arisen
from supernova precursors.  If so, then since the fast particles are C
and O, then there is a natural explanation of the observed correlation
of Be,B $\propto$ O in Pop II stars, i.e.\ a linear slope (rather than
the quadratic slope naively expected {}from galactic cosmic ray
production).  This possibility is intriguing and demands further
investigation (Vangioni-Flam et al.\ \cite{elisa}).

Regardless of the impact of our LiBeB yields on galactic chemical
evolution, the production is still high enough to create large local
enhancements of these elements.  This is particularly important if the
fast particles are well-confined by local magnetic fields and so the
yields are concentrated in a small volume.  Indeed, if sufficiently
localized, the fast particle yields would dominate the local LiBeB
abundances {\it and ratios}.  This possibility is interesting in light
of the Lemoine, Ferlet, \& Vidal--Madjar (\cite{martin}) observation
of a very localized Li isotopic anomaly in the direction of $\zeta$
Oph.  The observed \li7/\li6 ratio is $\sim 2 $ in one of the two
observed components; this is tantalizingly close to the high \emax
values for this ratio found in figure \ref{fig:lrat}.  If this is so,
then the other elemental and isotopic ratios in this region should be
as given in the figure; in fact, none of these are very different from
the solar ratios.  Thus the signature of supernova precursor
production would be for the other ratios to appear solar, and of
course, for the \li7/\li6 measurement to hold up.

Finally, if the process we discuss does produce significant Pop II
\li6, Be, and B, then one consequence is apparently that there are at
least two other important sources of these.  Galactic cosmic rays are
an inevitable and significant source; furthermore, their inability to
reproduce the
\b11/\b10 ratio suggests the need for another mechanism as well.
Perhaps the $\nu$-process is the remaining source, perhaps some other
low-energy particle spectrum (Cass\'e, Lehoucq, \& Vangioni-Flam
\cite{clv}; Ramaty, Kozlovsky, \& Lingenfelter \cite{rkla}),
or perhaps something else entirely.  At any rate, in explaining LiBeB
production and evolution, one seems to be faced at the moment with an
embarrassment of riches--more theories exist than problems to solve.
To resolve the issue will require not just more observational data but
also more accurate theory.

\section{Discussion}

We find that the prompt ejecta from a supernova may be behind the
curious $\gamma$-ray observations of Orion.  This mechanism has the
advantage of providing a definitive physical mechanism for a
low-energy particle spectrum with a non-solar composition, as seems
required by the observations.  To explain the observation we require
the progenitor of the responsible supernova to have been not a
``normal,'' type II supernova with a small wind, but instead an
exploding CO core star.  This leads not only to the observed
superabundance of low energy CO nuclei, but also predicts that there
will be
\ne22 (rather than \ne20) lines just below the current upper limits.
Consequently, our hypothesis is testable.

Furthermore, via the analytical supernova shock models of Nadyozhin
(\cite{nad}), we see that this mechanism is generic to supernova
explosions.  There are always precursor particles preceding the bulk
of the shock, and these particles inevitably have nuclear interactions
with their environment if \emax\ is sufficiently large.  The $\gamma$
production could be a new probe of the supernova explosion, providing
a detailed signature of the outermost shells of material.  This could
in principle be used as an independent means of determining supernova
masses (Fields, Cass\'e, Nomoto, Schramm, \& Vangioni-Flam
\cite{snsig}).

The implications for LiBeB evolution are unclear at the moment and
bear further investigation.  While it is not clear that the LiBeB
yields will play a major role in galactic chemical evolution, they
nevertheless could account for large local variations in these
elements.  Thus this process could offer a potential explanation for
the low $\li7/\li6$ ratio observed by Lemoine et al.\ (\cite{martin})
in the direction of $\zeta$ Oph.

We caution that calculations we present here are rough.  Some
improvements can be implemented without too much difficulty.  For
example, we have treated the propagation of the flux in a sketchy
manner, but as we indicate the full time dependence should be
considered.  Also, the flux first encounters the supernova
progenitor's wind, and the effect of traversing the wind should be
included.  Both of these effects will be addressed in a future paper
(Fields \cite{fcv}).

Another uncertainty in the calculations arises from the lack of
accurate cross section measurements for $\gamma$-ray and spallation
processes occurring near threshold.  We urge experimentalists to
examine this behavior.  Finally, there are some uncertainties to our
calculations which are more difficult to address accurately.  The most
notable of these is the effect of a magnetic field on the propagation
and stopping of the particles.  A sizable field could for example lead
to anisotropic emission of particles and so of $\gamma$-rays.
Spiraling in magnetic fields will also shorten the total displacement
of the fast particles from the supernova explosion site.  These
effects are however difficult to estimate.

Despite these problems, the precursor particles clearly have the
potential to provide a new probe of supernova explosions.  More work
on this mechanism is in order.

\bigskip

We thank Dave Schramm, Jim Truran, Don Ellison, and Simon Swordy for
stimulating discussions, and Robert Mochkovitch for sagacious advice
and a careful reading of the manuscript.  We especially thank Reuven
Ramaty for pointing out the \ti44 constraint.  This material is based
upon work supported by the North Atlantic Treaty Organization under a
Grant awarded in 1994.  EV-F and MC are supported in part by PICS
n$^\circ$114 of the CNRS (Origin and Evolution of the Light Elements).

Since this work was originally submitted, we became aware of the {\it
Letter} of Cameron et al.  (\cite{cam}), which suggests a type Ib
supernova as a source of the fast particles in Orion.

\clearpage

\begin{table}[htb]
\label{tab:shell}
\caption{Composition of outermost shells, by number}
\begin{center}
\begin{tabular}{ccccccc}
\hline\hline
Iniitial Mass & CO core mass & Ejected mass 
    & He & \c12 & \o16 & \ne22 \\
\hline
13 \msol & 1.8 \msol & 0.54 \msol & 1.44 &  1.39 & $\equiv$ 1 & 0.0203 \\
15 \msol & 2.1 \msol & 0.86 \msol & 1.61 &  1.38 & $\equiv$ 1 & 0.0337 \\
18 \msol & 2.9 \msol & 1.54 \msol &  684 &  6.42 & $\equiv$ 1 & 1.26 \\
\hline
\end{tabular}
\end{center}
\end{table}

\clearpage

\centerline{FIGURE CAPTIONS}

\bigskip
\bigskip

\begin{enumerate}

\item 
\label{fig:coflux}
The spectrum of the ejecta from a $2.1 \ \msol$ C+O core, for a
progenitor originally of mass 15 \msol.  Note the absence of protons
and the dominance of C and O over $\alpha$ particles.  The small
bounciness is due to numerical inaccuracy, but the large jumps, e.g.\
in the C spectrum, are real and arise from changes in the compositions
of the various nuclear burning shells.  In the energy range of
interest ($E \ga 1$ MeV/nucl), the spectrum is well approximated by a
power law with a slope of $\sim 4$, in good agreement with the
analytic results by Nadyozhin (\cite{nad}).  The endpoints of the
curves at $E \sim 3.5 MeV$ are artifacts of the supernova model and do
not correspond to \emax.

\item 
\label{fig:emiss}
(a) $\c12^*$, $\o16^*$, and $\ne22^*$ emission as a function of \emax.
Curves are normalized so that the total CO $\gamma$ emission matches
the Bloemen et al.\ (\cite{blo}) Orion observations, i.e.,
$\dot{N}({\rm C}^*+{\rm O}^*) = 3\times 10^{39} \ {\rm s}^{-1}$.  This
amounts to fixing the hydrogen density $n_{\rm H}$ at each \emax. \\
(b) The required hydrogen density $n_{\rm H}$ as a function of \emax.
Note that the density is fairly constant over \emax, and is only
modestly above the average Orion density.  The rise at low \emax\
stems from the lower particle flux above $\gamma$-ray production
energies.

\item
\label{fig:escal}
(a) The spectrum (as in figure \ref{fig:coflux}) for explosion
energies $U = 0.6 \times 10^{51}$, $0.8 \times 10^{51}$, and $1 \times
10^{51}$ erg, all with CO core mass $2.1 \msol$.  \\
(b) As in (a), with the energies rescaled according to the rule
$E^* = (U/10^{51} \ {\rm erg})^{-1} E$; 
for the analytic scaling law this should give
identical spectra.  While this is not
perfectly true, clearly the scaling is a good approximation.

\item
\label{fig:masscal}
(a) The spectrum (as in figure \ref{fig:coflux}) for CO core masses
$1.8$, $2.1$, and $2.9 \msol$, all having an explosion energy of $U =
1 \times 10^{51}$ erg.  Although the shapes are similar, the higher
energy scale for the lower masses means that they also have larger
intensities in the energies above threshold. \\
(b) As in (a), with the energies rescaled according to the rule $E^* =
(M/2.1 \msol) E$; for the analytic scaling law this should give
identical spectra and the curves should overlap.  We see that this
scaling is not very good, and the results seen more consistent with
little or no mass dependence.

\item
\label{fig:mout}
The yields of LiBeB isotopes
as a function of the cutoff energy \emax.  
The strong \emax\ dependence is a result of crossing the
various thresholds for spallation production.
Note that the yields are very small, even though our 
CO core supernova must be considered as a best (and uncommon)
case.  Thus it seems unlikely that the LiBeB production by 
fast ejecta is significant for chemical evolution.

\item
\label{fig:lrat}
Elemental and isotopic ratios of LiBeB as a function of \emax.
The extreme behavior at low \emax\ is due to threshold behavior.
Since the absolute yields are small, these ratios are only of
interest if they may be observed in a very localized region.
The strong variations at low \emax\ are due to cross section
threshold and resonance effects.

\end{enumerate}


\clearpage

\begin{thebibliography}

\bibitem[1989]{ag} {Anders, E., \& Grevesse, N. 1989, Geochim.\ et
Cosmochim.\ Acta, 53, 197}

\bibitem[1987]{be} {Blanford, R., \& Eichler, D., 1987, 
Phys.\ Rep., 154, 1}

\bibitem[1993]{blo} {Bloemen, H., et al.\ 1993, A\&A, 281, L5}

\bibitem[1991]{brown} {Brown, L.E., Dearborn, D.S., Schramm, D.N.,
Larsen, J.T., \& Kurokawa, S. 1991, ApJ, 371, 648}

\bibitem[1994]{bb} {Bykov, A., \& Bloemen, H. 1994, A\&A, 283, L1}

\bibitem[1995]{cam} {Cameron, A.G.W., H\"oflich, P.,
Meyers, P.C. \& Clayton, D.D. 1995, ApJ, 447, L53}

\bibitem[1996]{snsig}{Cass\'e, M., Fields, B.D., Nomoto, K.,
Schramm, D.N., \& Vangioni-Flam, E. 1996 in preparation}

\bibitem[1978]{cg}{Cass\'e, M., \& Goret, H. 1978, ApJ, 221, 703}

\bibitem[1995]{clv} {Cass\'e, M., Lehoucq, R., \& Vangioni-Flam, E.
1995, Nature, 373, 318}

\bibitem[1994]{cvlo} {Cass\'e, M., Vangioni-Flam, E., Lehoucq, R., \& 
Oberto, Y. 1994, in Light Element Abundances, eds. P. Crane
\& J. Faulkner, 389}

\bibitem[1994]{cbhw} {Clocchiatti, A., Brotherton, M., Harkness, R.P., \&
Wheeler, J.C. 1994, IAU Circ. No.\ 5972}

\bibitem[1973]{colg} {Colgate, S.A. 1973, ApJ, 181, L53}

\bibitem[1960]{cj} {Colgate, S.A., \& Johnson, M.H. 1960, Phys.\ Rev.\
Lett., 5, 235}

\bibitem[1989]{dear} {Dearborn, D., Schramm, D.N., Steigman, G.,
\& Truran, J. 1989, ApJ, 347, 455}

\bibitem[1995]{dhm} {Digel, S.W., Hunter, S.D., Mukherjee, R. 
1995, ApJ, 441, 270}

\bibitem[1992]{eb} {Ensman, L., \& Burrows, A. 1992, ApJ, 393, 742}

\bibitem[1991]{ell} {Jones, F.C., \& Ellison, D.C. 1991, Sp.\ Sci.\ Rev.,
58, 259}

\bibitem[1996]{fcv} {Fields, B.D. 1996 in preparation}

\bibitem[1990]{gais} {Gaisser, T.K. 1995, Cosmic Rays and Particle Physics,
(Cambridge: Cambridge University Press), 149}

\bibitem[1982]{goudis} {Goudis, C. 1982, The Orion Complex: A Case Study of
Interstellar Matter (D. Ridel: Dodrecht), 12}

\bibitem[1994]{iwa} {Iwamoto, K., et al.\ 1994, ApJ, 437, 115}

\bibitem[1994]{ti44} {Iyudin et al.\ 1994, A\&A, 284, L1}

\bibitem[1995]{jc} {Jin, L., \& Clayton, D.D. 1995, ApJ, 447, L53}

\bibitem[1991]{kum} {Kumagai, S., Shigeyama, T., Hashimoto, M., \& 
Nomoto, K. 1991, A\&A, 243, L113}

\bibitem[1993]{martin} {Lemoine, M., Ferlet, R., Vidal--Madjar, A. 
1995, A\&A, in press}

\bibitem[1975a]{mra} {Meneguzzi, M, \& Reeves, H. 1975a, A\&A, 40, 91}

\bibitem[1975b]{mrb} {Meneguzzi, M, \& Reeves, H. 1975b, A\&A, 40, 99}

\bibitem[1985a]{meya} {Meyer, J.-P. 1985a, ApJS, 57, 151}

\bibitem[1985b]{meyb} {Meyer, J.-P. 1975b, ApJS, 57, 173}

\bibitem[1977]{montm} {Montmerle, T. 1977, ApJ, 217, 872}

\bibitem[1994]{nad} {Nadyozhin, D.K. 1994, in Les Houches, Session LIV 1990:
Supernovae, ed.\ S. Bludman, R. Mochkovitch, \& J. Zinn-Justin
(Elsevier: New York),569}

\bibitem[1994]{nb} {Nath, B., \& Biermann, P. 1994, MNRAS, 270, L33}

\bibitem[1991]{nom91} {Nomoto, K., Kumagai, S., \& Shigeyama, T. 1991,
in Gamma-Ray Line Astrophysics, eds., P. Durouchoux \& N. Prantzos
(AIP 232: New York), 236}

\bibitem[1994]{nomoto} {Nomoto, K., et al.\ 1994, Nature, 471, 227}

\bibitem[1995a]{nom95a} {Nomoto, K., et al.\ 1995a, in IAU
Symposium 165, Compact Stars in Binaries, ed.\ J. van Paradij
\& E.P.J. van den Heuvel (Kluwer:  Dodrecht), in press}

\bibitem[1995b]{nom95b} {Nomoto, K., et al.\ 1995b, 
in 17th Texas Symposium on Relativistic Astrophyiscs (Ann.\ N.Y.
Acac.\ Sci.: New York), in press}

\bibitem[1994]{opsv} {Olive, K.A., Prantzos, N., Scully, S., \& 
Vangioni-Flam, E. 1994, ApJ, 424, 666}

\bibitem[1979]{rkl79} {Ramaty, R., Kozlovsky, B, \& Lingenfelter, R.E. 
1979, ApJS, 40, 487}

\bibitem[1995a]{rkla} {Ramaty, R., Kozlovsky, B, \& Lingenfelter, R.E. 
1995a, ApJ, 438, L21}

\bibitem[1995b]{rklb} {Ramaty, R., Kozlovsky, B, \& Lingenfelter, R.E. 
1995b, ApJ, 438, L21}

\bibitem[1996]{rvfc} {Ramaty, R., Vangioni-Flam, E., Fields, B.D.,
\& Cass\'e, M. 1996, in preparation}

\bibitem[1984]{rv} {Read, S.M., \& Viola, V.E. 1984, Atom.\ Data
\& Nuc.\ Data Tables, 31, 359}

\bibitem[1995]{rp} {Reeves, H., \& Prantzos, N. 1995, in The Light
Element Abundances, eds.\ P. Crane \& J. Faulkner, 382}

\bibitem[1994]{sck} {Schmidt, B., Challis, P., \& Kirchner, R.
1994, IAU Circ. No.\ 5966}

\bibitem[1992]{sw} {Steigman, G., \& Walker, T.P. 1992, ApJ, 385, L13}

\bibitem[1994]{f-kt94} {Thielemann, F.-K., Nomoto, K., \& Hashimoto,
M., 1994, in Supernovae, Les Houches, Session LIV, eds.\ S. Bludman,
R. Mochkovitch, \& J. Zinn-Justin (Elsevier: Amsterdam), 629}

\bibitem[1996]{f-kt96} {Thielemann, F.-K., Nomoto, K., \& Hashimoto,
M. 1996, ApJ, in press}

\bibitem[1995]{tww} {Timmes, F.X., Woosley, S.E., \& Weaver, T.A.
1995, ApJ, in press}

\bibitem[1996]{elisa} {Vangioni-Flam, E., Olive, K.A., Fields, B.D.,
\& Cass\'e, M. 1996, ApJ sumitted}

\bibitem[1985]{wmv} {Walker, T.P., Mathews, G.J., \& Viola, V.E.
1985, ApJ, 299, 745}

\bibitem[1976]{weaver} {Weaver, T., 1976, ApJS, 32, 233}

\bibitem[1993]{wlw} {Woosley, S.E., Langer, N.
\& Weaver, T.A. 1993, ApJ, 411, 823}

\bibitem[1995]{ww} {Woosley, S.E., \& Weaver, T.A. 1995, ApJ, in preparation}

\end{thebibliography}
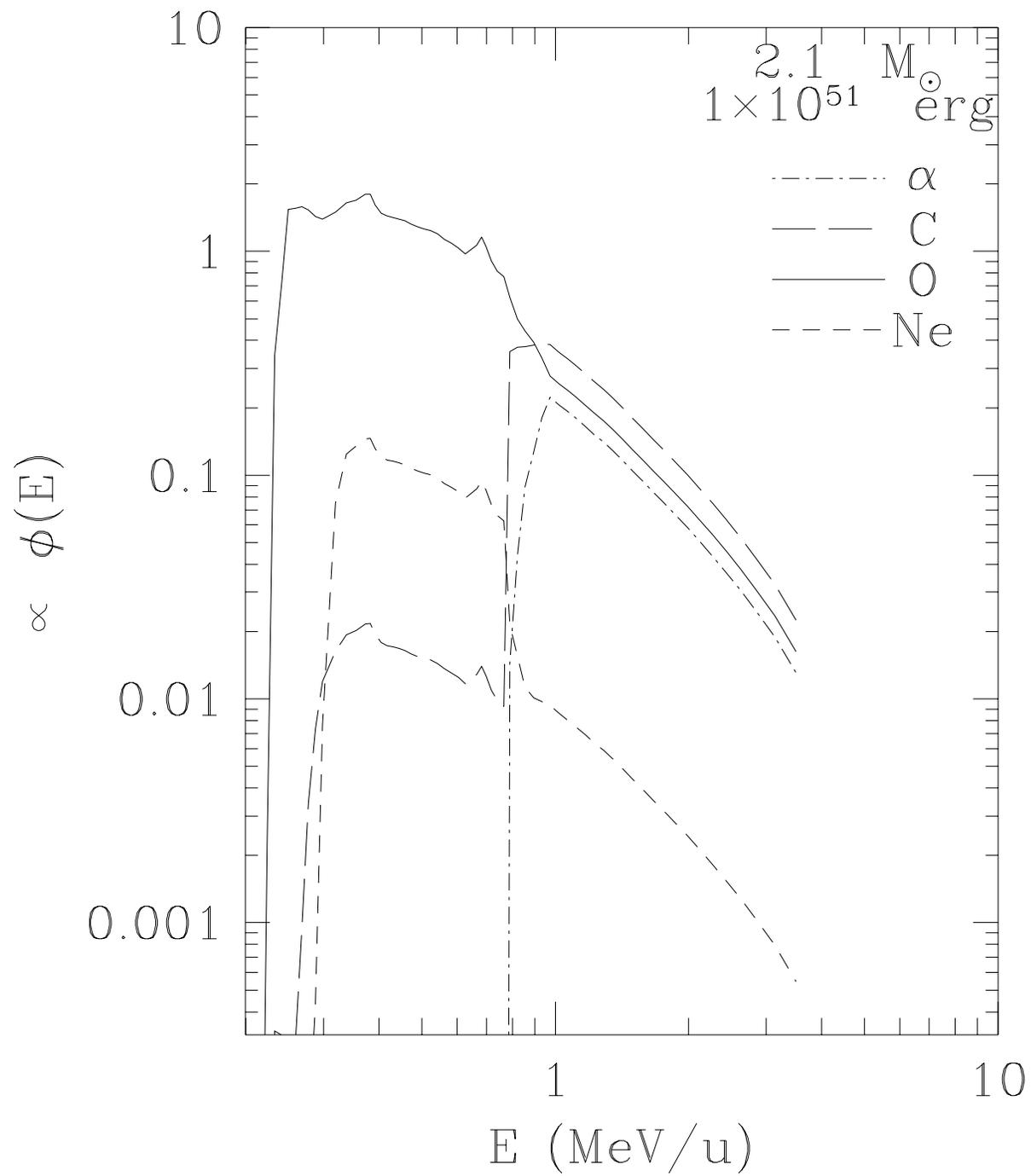
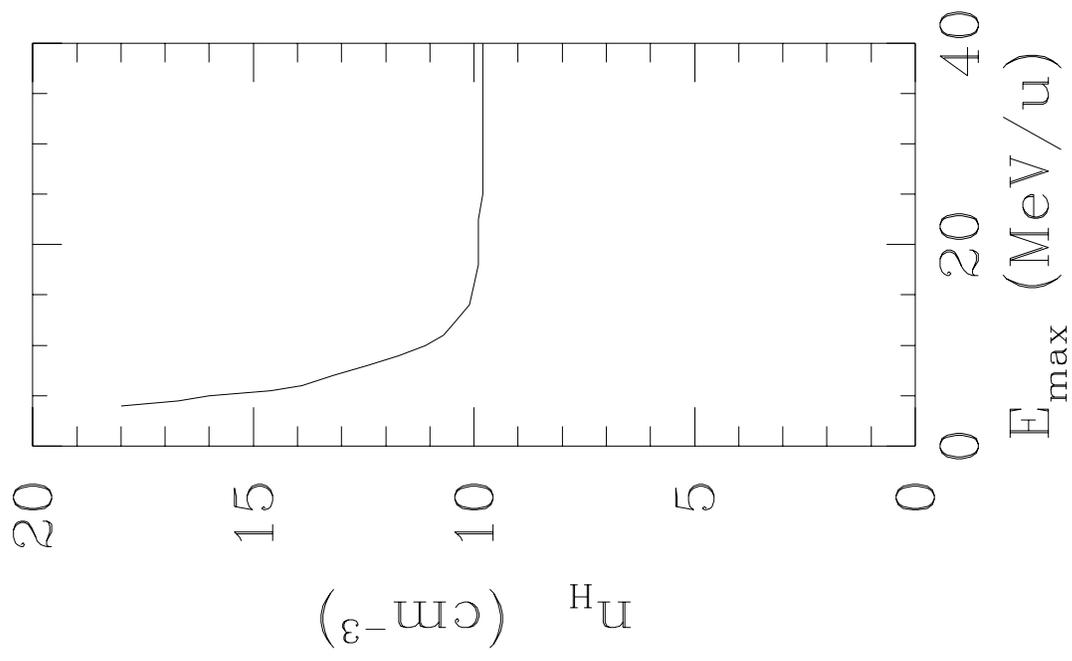
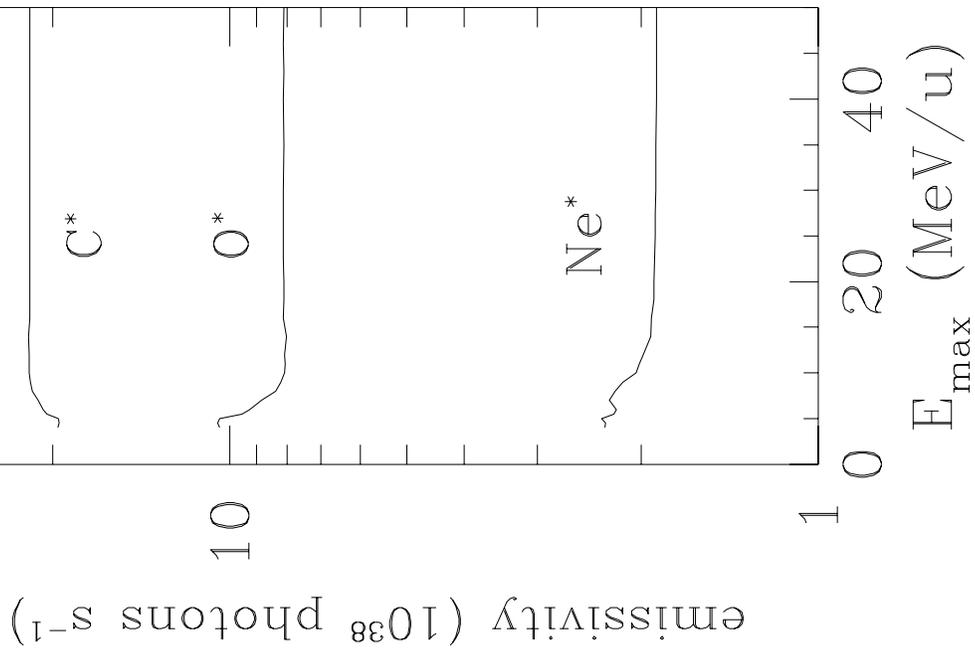
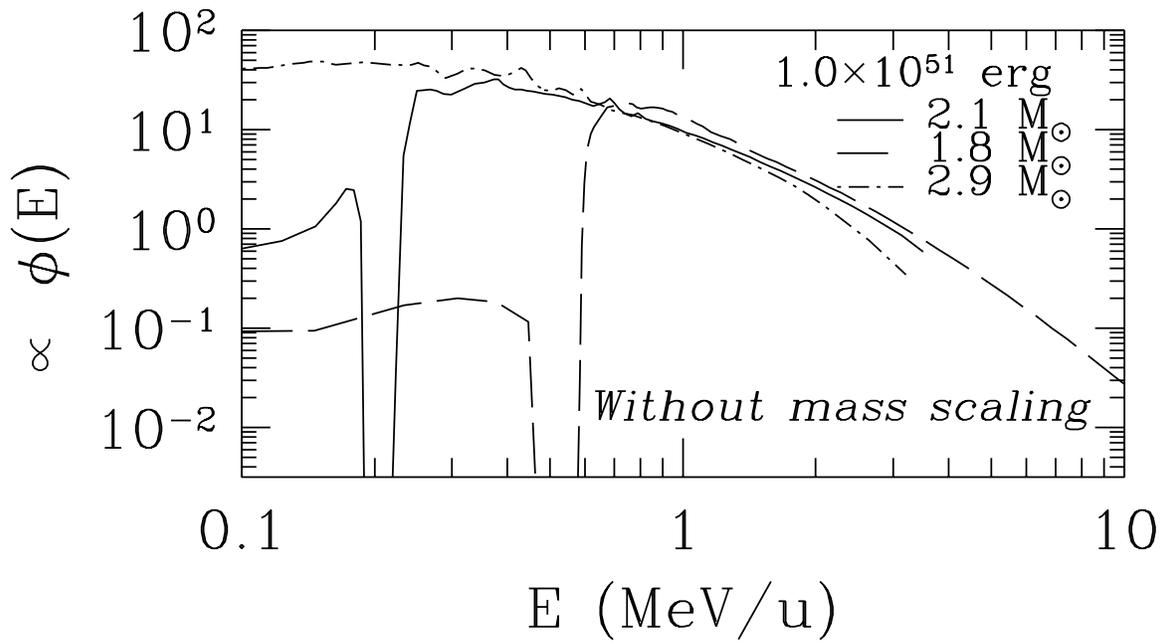
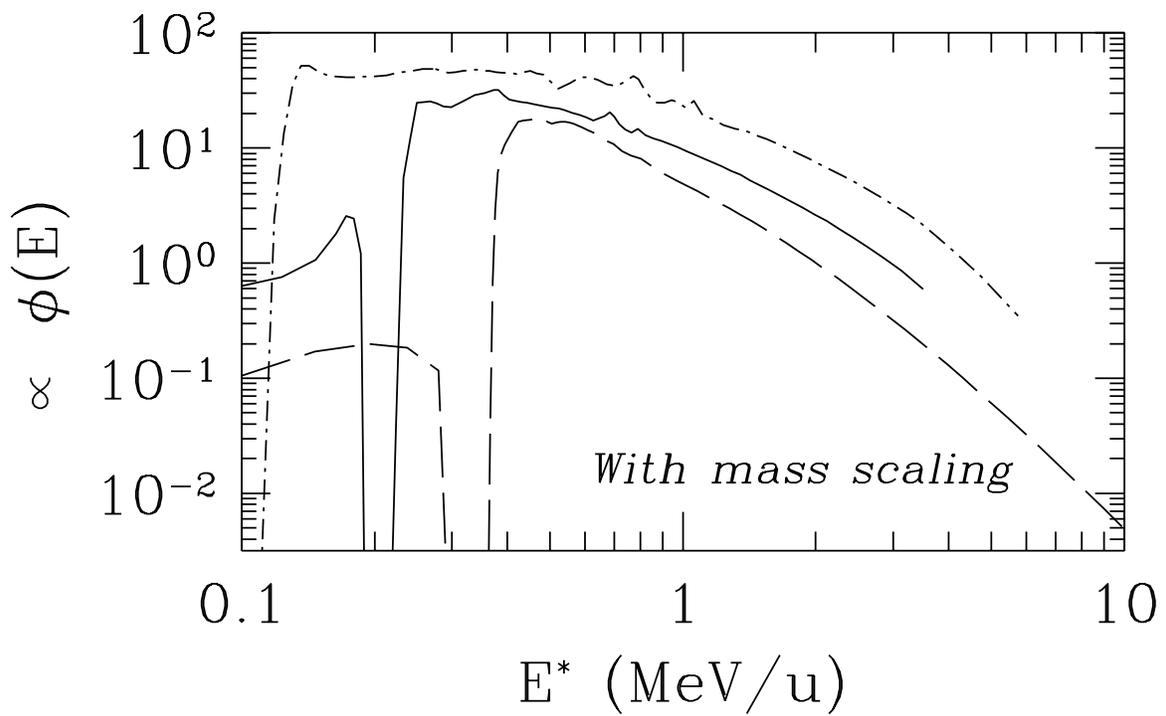
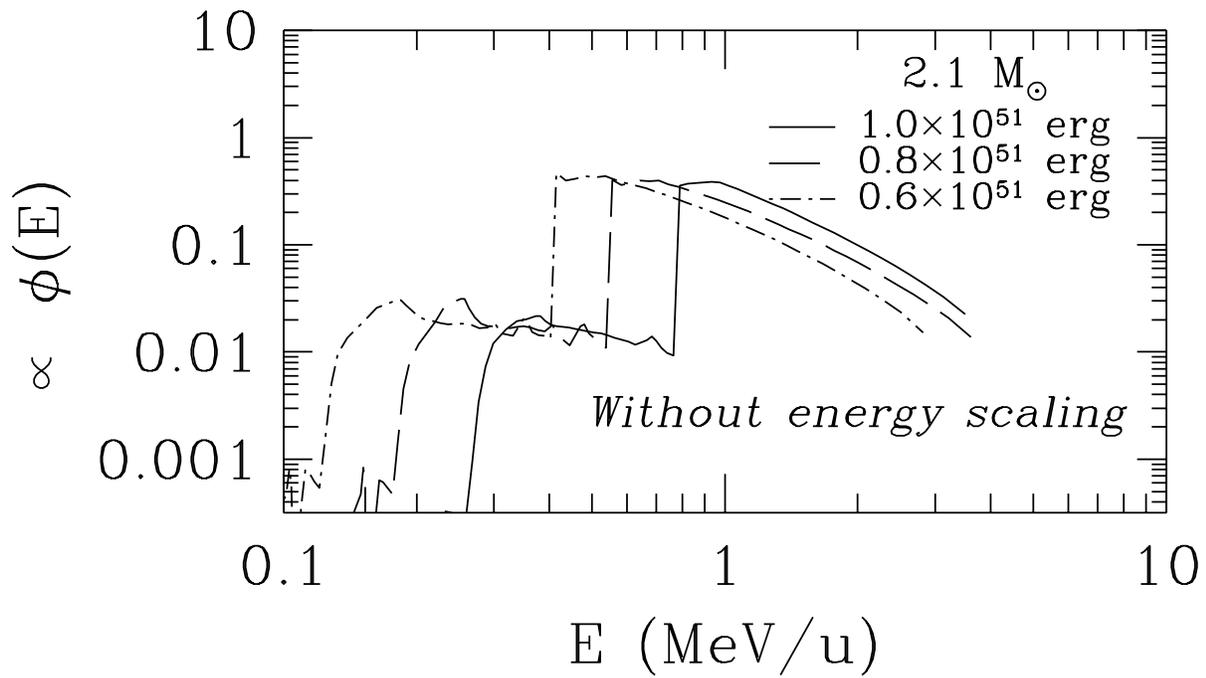
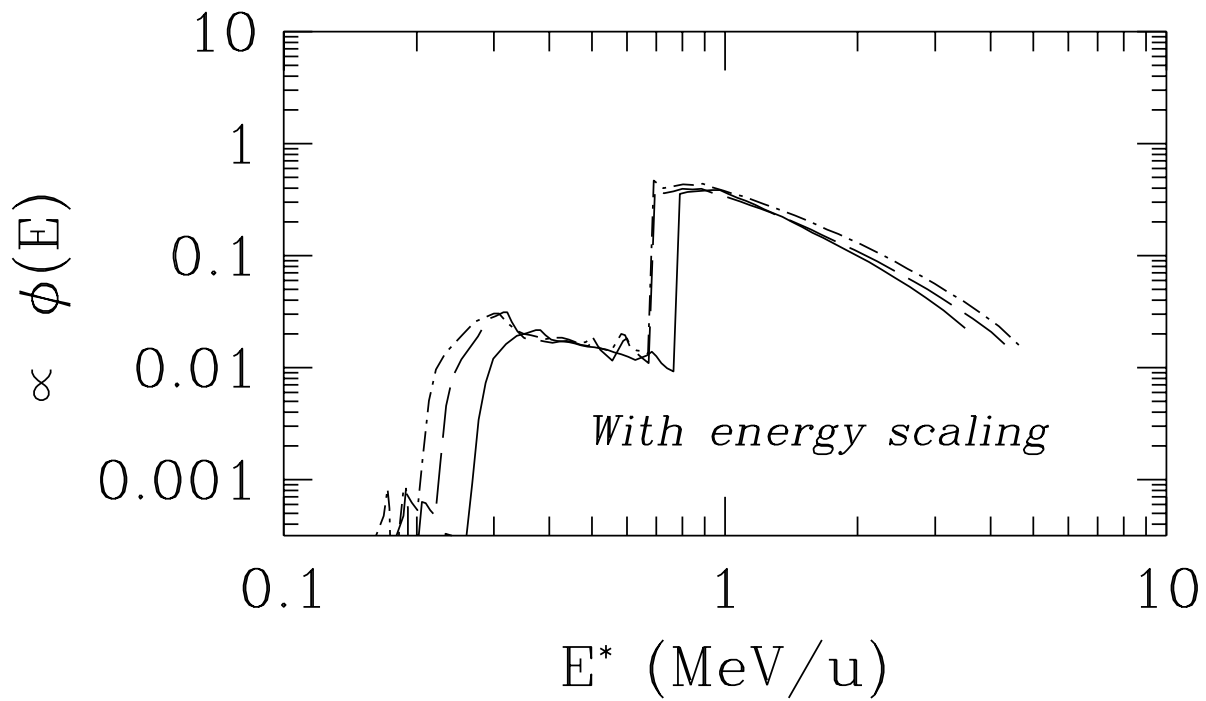
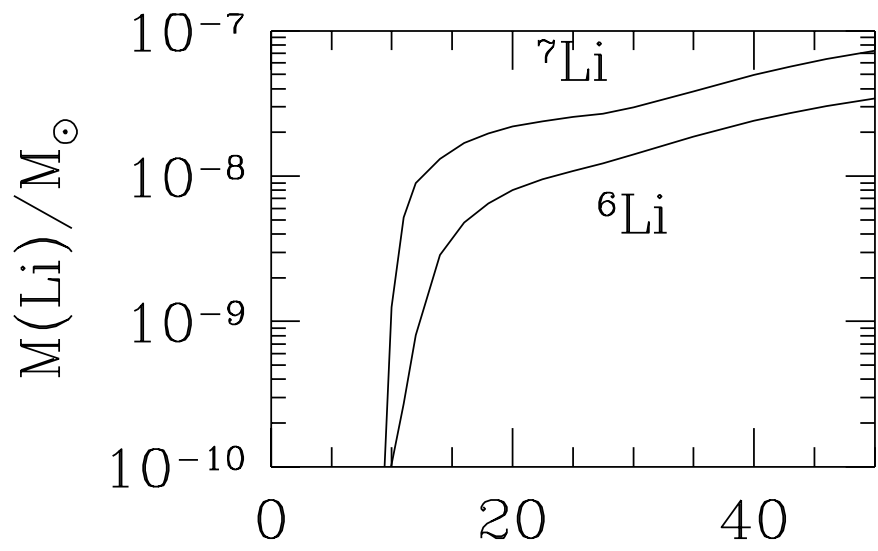
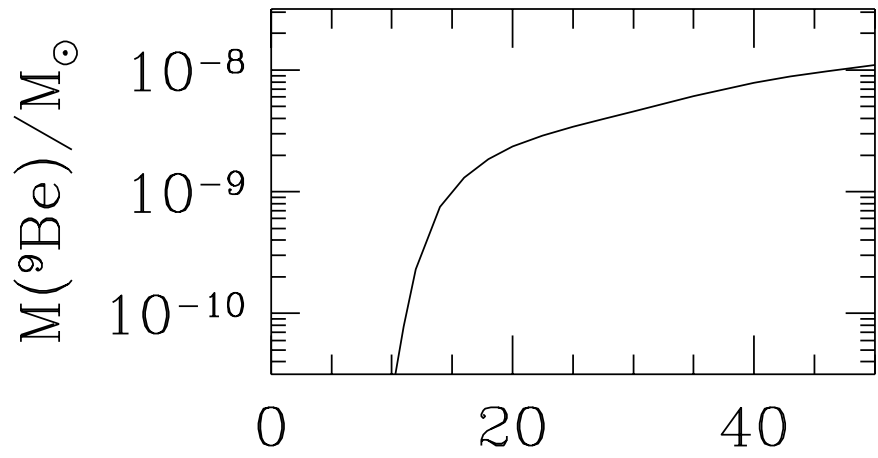
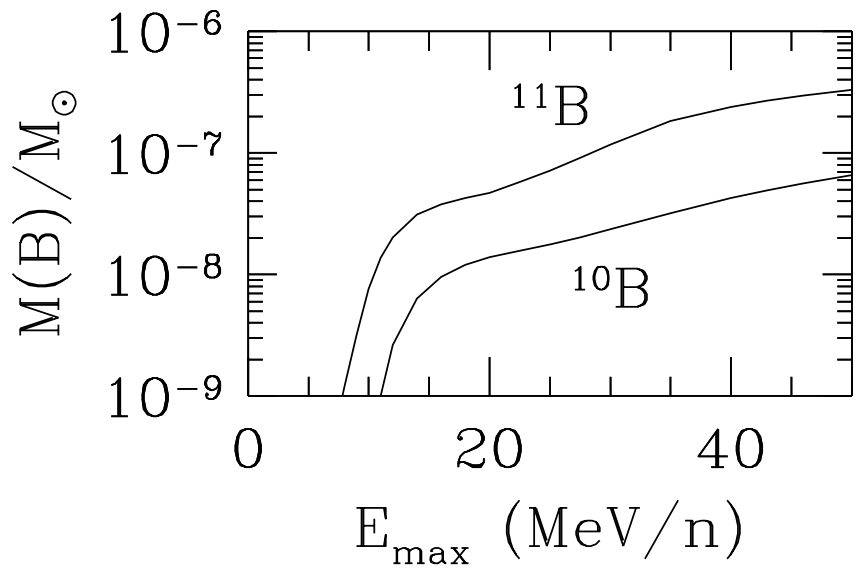
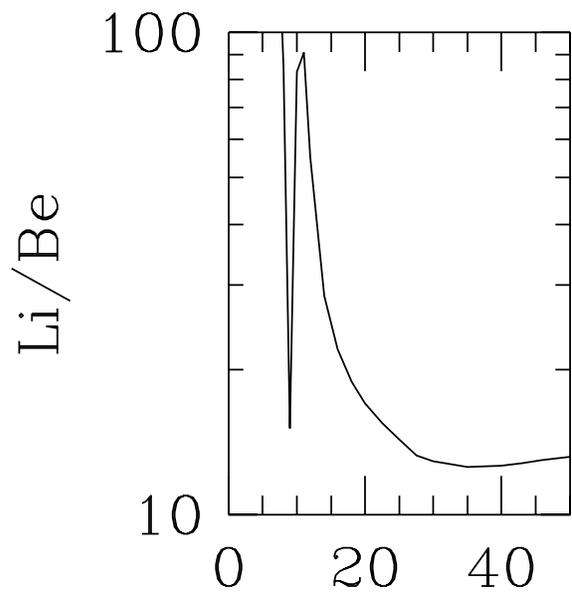
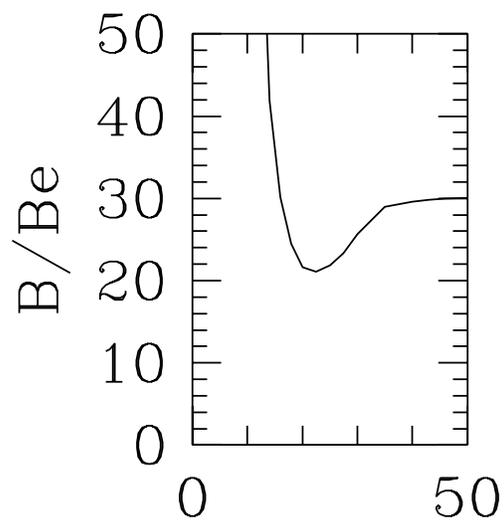
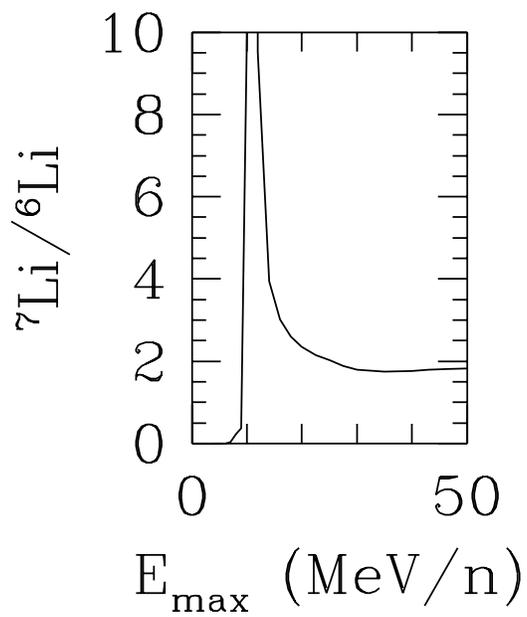
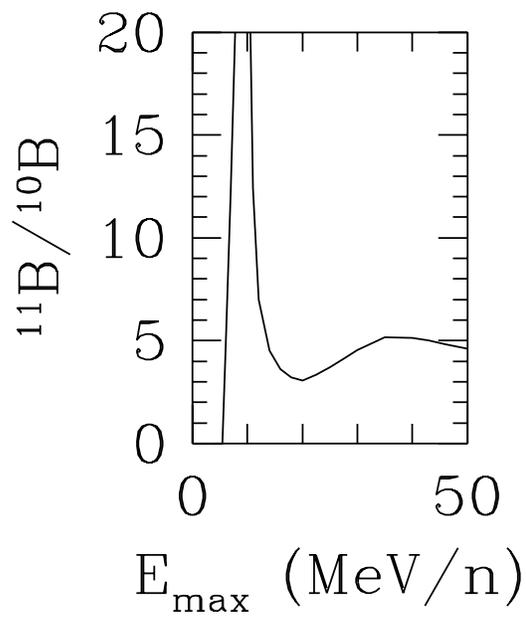
\end{document}